\documentclass[preprint,pteplogo,amsmath,amssymb,aps]{ptephy}

\preprintnumber{XXXX-XXXX}

\usepackage{graphicx} 
\usepackage{dcolumn}  
\usepackage{bm}       

\usepackage{hyperref}

\begin{document}

\title{Improved {\bf accuracy in} determination of thermal cross section of ${}^{14}{\rm N}({\rm n},{\rm p}){}^{14}{\rm C}$ for the neutron lifetime measurement
}

\author{\name{R.~Kitahara}{1*}, \name{K.~Hirota}{2,3}, \name{S.~Ieki}{4,5}, \name{T.~Ino}{6,7}, \name{Y.~Iwashita}{8}, \name{M.~Kitaguchi}{3,9}, \name{J.~Koga}{10}, \name{K.~Mishima}{6,7**}, \name{A.~Morishita}{10}, \name{N.~Nagakura}{4}, \name{H.~Oide}{4,11}, \name{H.~Otono}{12}, \name{Y.~Seki}{7,13}, \name{D.~Sekiba}{14}, \name{T.~Shima}{2}, \name{H.~M.~Shimizu}{3}, \name{N.~Sumi}{10}, \name{H.~Sumino}{15},\name{K.~Taketani}{6,7}, \name{T.~Tomita}{10}, \name{T.~Yamada}{4}, \name{S.~Yamashita}{16}, \name{M.~Yokohashi}{3}, and \name{T.~Yoshioka}{12}
}

\address{
\affil{1}{Department of Physics, Kyoto University, Kitashirakawa, Kyoto, 606-8502, Japan}
\affil{2}{Research Center for Nuclear Physics (RCNP), Osaka University, Ibaraki, Osaka 567-0047, Japan}
\affil{3}{Department of Physics, Nagoya University, Chikusa, Nagoya 464-8602, Japan}
\affil{4}{Department of Physics, The University of Tokyo, Bunkyo, Tokyo 113-0033, Japan}
\affil{5}{Research Centre for Neutrino Science, Tohoku University, Sendai, Miyagi, 980-8578, Japan}
\affil{6}{High Energy Accelerator Research Organization (KEK), Tsukuba, Ibaraki 305-0802, Japan}
\affil{7}{J-PARC Center, Tokai, Ibaraki 319-1195, Japan}
\affil{8}{Institute of Chemical Research, Kyoto University, Uji, Kyoto 611-0011, Japan}	
\affil{9}{Kobayashi-Maskawa Institute for the Origin of Particles and the Universe (KMI), Nagoya University, Chikusa, Nagoya 464-8602, Japan}
\affil{10}{Department of Physics, Graduate School of Science, Kyushu University, Fukuoka, Fukuoka 819-0395, Japan}
\affil{11}{now at Department of Physics, Tokyo Institute of Technology, Meguro, Tokyo 152-8550, Japan}
\affil{12}{Research Center for Advanced Particle Physics (RCAPP), Kyushu University, Fukuoka, Fukuoka 819-0395, Japan}
\affil{13}{Japan Atomic Energy Agency, Tokai, Ibaraki, 319-1195, Japan}
\affil{14}{Institute of Applied Physics, University of Tsukuba, Tennoudai 1-1, Tsukuba, Ibaraki 305-8573, Japan}
\affil{15}{Department of Basic Science, Graduate School of Arts and Sciences, The University of Tokyo, Meguro, Tokyo 153-8902, Japan}
\affil{16}{International Center for the Elementary Particle Physics (ICEPP), The University of Tokyo, Bunkyo, Tokyo 113-0033, Japan}
\email{kitahara@kyticr.kuicr.kyoto-u.ac.jp, kenji.mishima@kek.jp}
}
\date{\today}
\begin{abstract}
In a neutron lifetime measurement at the Japan Proton Accelerator Complex, the neutron lifetime is calculated by the neutron decay rate and the incident neutron flux.
The flux is obtained due to counting the protons emitted from the neutron absorption reaction of ${}^{3}{\rm He}$ gas, which is diluted in a mixture of working gas in a detector.
Hence, it is crucial to determine the amount of ${}^{3}{\rm He}$ in the mixture.
In order to improve the accuracy of the number density of the ${}^{3}{\rm He}$ nuclei, we suggested to use the ${}^{14}{\rm N}({\rm n},{\rm p}){}^{14}{\rm C}$ reaction as a reference because this reaction involves similar kinetic energy as the $^3$He(n,p)$^3$H reaction and a smaller reaction cross section to introduce reasonable large partial pressure.
The uncertainty of the recommended value of the cross section, however, is not satisfied with our requirement.

In this paper, we report the most accurate experimental value of the cross section of the $^{14}$N(n,p)$^{14}$C reaction at a neutron velocity of 2200 m/s, measured relative to the $^3$He(n,p)$^3$H reaction. 
The result was 1.868 $\pm$ 0.003 (stat.) $\pm$ 0.006 (sys.) b.
Additionally, the cross section of the $^{17}$O(n,$\alpha$)$^{14}$C reaction at the neutron velocity is also redetermined as 249 $\pm$ 6 mb.
\end{abstract}
\subjectindex{C30, D23, H11}
\maketitle
\section{INTRODUCTION}
\hspace{\parindent}
The neutron lifetime, $\tau_{\rm n}$, is an important parameter in cosmology and particle physics~\cite{tanabashi2018PDG}.
According to the Big Bang nucleosynthesis, the light nuclei such as helium and lithium were formed in the early universe due to the collisions between protons and neutrons, and $\tau_{\rm n}$ can be used to predict the abundance of these light nuclei in the universe. 
The Cabibbo--Kobayashi--Maskawa matrix element, $V_{ud}$, is also calculated using $\tau_{\rm n}$ because the neutron decays are caused by weak interactions.

At present, there are two prevalent methods for measuring the neutron lifetime; the storage method~\cite{Mampe1993an, pichlmaier2010neutron, steyerl2012quasielastic, Arzumanov2015tea, serebrov2017new, pattie2018measurement, ezhov2018measurement} and the beam method~\cite{Byrne1996zz,yue2013improved}.
The neutron lifetime measured by the storage method yields a value of $879.4 \pm 0.4$ s, and the beam method evaluates the lifetime as $888.0 \pm 2.0$ s.
The recommended value for the neutron lifetime is $\tau_{\rm n}=880.2 \pm 1.0$ s~\cite{tanabashi2018PDG}.
The difference in the measured lifetime values from different techniques suggests a method-dependent discrepancy, which reduces the reliability of these measurements.
In addition to the method-related uncertainties, undiscovered decays may also cause discrepancy in the measurements~\cite{serebrov2008experimental,fornal2018dark}.
Hence, an appropriate evaluation of each experimental method to pin-point the major issues causing this discrepancy in measuring $\tau_{\rm n}$ is required.
The storage method has measurement uncertainty below 0.1\%, and thus new measurements by the beam method with similar accuracies are eagerly awaited.

Generally, the beam method yields $\tau_{\rm n}$ from the measured emission rate of decay particles and the incident neutron flux. In the latest beam-method experiment in which protons from neutron decay were detected using a proton trap, the neutron flux was measured with a neutron monitor that counted the neutron-induced charged particles from the $^6$Li(n,$\alpha$)$^3$H reaction~\cite{yue2013improved}.
Currently, another experiment based on the beam method is in progress at the Japan Proton Accelerator Research Complex (J-PARC)~\cite{nagakura2017precise}.
In this experiment, a time projection chamber (TPC) can simultaneously detect the electrons from the neutron decays as well as the protons emitted in the neutron absorption reactions due to the dilution of ${}^{3}{\rm He}$ in the gas mixture contained in the TPC.
The use of a single detector to measure both the decay rate and the incident neutron flux is useful in determining $\tau_{\rm n}$ with different systematic uncertainty as compared to other experimental methods.
The lifetime, $\tau_{\rm n}$ is calculated using the relation shown in
Eq.~\ref{eqtau}.
\begin{eqnarray}
\label{eqtau}
\tau_{\rm n} = \frac{1}{\rho({}^{3}{\rm He}) \sigma({}^{3}{\rm He}) v_0} \left(
\frac{S({}^{3}{\rm He})/\varepsilon({}^{3}{\rm He})}{S(\beta)/\varepsilon(\beta)}
\right) ,
\end{eqnarray}
where $S( ^{3} {\rm He})$ and $S({\beta})$ are the number of events related to the $^3$He(n,p)$^3$H and the neutron decay reactions, respectively; $\varepsilon(^{3}{\rm He})$ and $\varepsilon({\beta})$ are the respective efficiencies of each reaction; $\rho(^{3}{\mathrm He})$ is the number density of the $^3$He nuclei in a sealed vacuum vessel (TPC vessel); $\sigma(^{3} {\rm He})= 5333 \pm 7$ b~\cite{Mughabghab2006} is the cross section of the $^3$He(n,p)$^3$H reaction at the neutron velocity of $v_0=2200$ m/s.
When the statistical error, which is the major source of uncertainty currently, is suppressed, the systematic uncertainty of $\rho({}^{3}{\rm He})$ is expected to become dominant.
In this work, our approach is to minimize the systematic uncertainty associated with the neutron lifetime measurement.
In our experimental set up, the partial pressure of ${}^{3}{\rm He}$ is adjusted to a reasonably low value of 100 mPa by mixing the isopure ${}^{3}{\rm He}$ and commercially provided high-grade helium gas (G1He) so that the proton rate is similar to the electron rate.
The amount of ${}^{3}{\rm He}$ from the isopure ${}^{3}{\rm He}$ gas (with purity greater than $99.95$\%) in the TPC vessel is controlled by a commercial pressure gauge with an accuracy of 0.1\%.
However, a small amount of ${}^{3}{\rm He}$ present in the G1He still remains a source of uncertainty in determining the ${}^{3}{\rm He}$ content in the TPC vessel accurately.

This uncertainty could be reduced by introducing a mixture of G1He and nitrogen gas in the TPC vessel for performing ${}^{3}{\rm He}$ content measurements.
Using the ratio of the count rate of protons from the ${}^{3}{\rm He}({\rm n},{\rm p}){}^{3}{\rm H}$ and ${}^{14}{\rm N}({\rm n},{\rm p}){}^{14}{\rm C}$ reactions, we can determine the ${}^{3}{\rm He}$ content in the G1He, relative to the nitrogen content.
This method can provide the same accuracy as the measurements of the cross sections in the ratio of $\sigma(^{14} {\rm N})/\sigma(^{3} {\rm He})$, where $\sigma(^{14} {\rm N})$ is the cross section of the ${}^{14}{\rm N}({\rm n},{\rm p}){}^{14}{\rm C}$ reaction at $v_0$.
However, the recommended value of $\sigma(^{14} {\rm N}) = 1.86 \pm 0.03$ b~\cite{Mughabghab2006} is not suitably accurate for the calculation of the lifetime using Eq.~\ref{eqtau} because it introduces an uncertainty of $0.2\%$, assuming that 10\% ${}^{3}{\rm He}$ in the TPC vessel is present filled from the G1He.
In this paper, we report an improved method for measurement of $\sigma(^{14} {\rm N})/\sigma(^{3} {\rm He})$ ratio using the TPC, based on the experiment proposed by Kii {\it et al.}~\cite{Kii1999}, which may be useful in reducing the uncertainties associated with the neutron lifetime measurement.
\section{PROCEDURE}
\label{MeasPrinc}
In the TPC, a mixture of 80-kPa G1He and 20-kPa N$_2$ gas was used to measure $\sigma(^{14} \mathrm{N})$.
The specific pressure of the isopure $^{3}$He gas (mixed in the TPC gas with the G1He and N$_2$) at two different values ($\sim$10 Pa) was selected to observe the same event rate as the $^3$He(n,p)$^3$H and $^{14}$N(n,p)$^{14}$C reactions.
For identification purposes, we have defined the $^{3}$He gas with the lower and the higher pressures as gas I and gas I\hspace{-.1em}I respectively, in this paper.
The cross section, $\sigma(^{14} {\rm N})$ can be obtained using Eq.~\ref{eqNsigma}.
\begin{eqnarray}
\label{eqNsigma}
\sigma(^{14} \mathrm{N})=\frac{\epsilon(^{3} \mathrm{He})} {\epsilon(^{14} \mathrm{N})}
\frac{S(^{14} \mathrm{N})}{S(^3 \mathrm{He})}
\frac{\rho(^3 \mathrm{He}) }{\rho(^{14} \mathrm{N})}
\sigma(^3 \mathrm{He}),
\end{eqnarray}
where $\varepsilon(^{14} \mathrm{N})$ is the detection efficiency and $S(^{14} \mathrm{N})$ is the number of events of the $^{14}$N(n,p)$^{14}$C reaction. 
The ratio, $\rho(^3 \mathrm{He})/\rho(^{14} \mathrm{N})$ was measured using a gas handling system.
It should be noted that $\rho(^{14} \mathrm{N})$ is measured with an uncertainty of 0.3\%, which has a better accuracy than the reported measurements for a solid target~\cite{14NWagemans2000}.

In our measurements, the bunched neutron beams that are shorter than a length of the TPC are used.
As the whole bunched-neutron beam enters the sensitive volume of the TPC, the reaction events are counted, and the total reaction energy is deposited in the TPC, because the ions and protons stop in this volume.
In addition, the measurements performed in the TPC have a low background environment due to the absence of other neutron absorption reactions inside the volume.
The deposited energy distribution is expected to have two narrow energy peaks at 0.764 MeV and 0.626 MeV corresponding to the deposited energy of the $^3$He(n,p)$^3$H and the $^{14}$N(n,p)$^{14}$C reactions, respectively.
\section{MEASUREMENT}
\label{MEASUREMENT}
\subsection{BUNCHED NEUTRON BEAM}
The pulsed neutron beams are produced by the spallation process at the Materials and Life Science Experimental Facility in J-PARC, in which a mercury target is irradiated with 3 GeV protons carrying a maximum current of 333 ${\rm \mu}$A and at a repetition rate of 25 Hz.
In the experimental beam line of BL05~\cite{Mishima2009}, the polarized beams with a polarization ratio of 94--97\%~\cite{Ino2011} were used for bunching neutrons to an arbitrary length using the spin-flip chopper (SFC) \cite{Taketani2011}.
At the end of the SFC, a neutron beam monitor \cite{Ino2014} was attached for monitoring the incident neutron flux.
To study the beam-independent background signal (i.e. caused by the cosmic rays), the measurement without the neutron beam was also performed. 
The injection of the neutron beam was controlled by a beam-switching shutter made of $^6$Li tiles~\cite{Arimoto2015}. 
To eliminate the double counting of events, the incident neutron flux was reduced by a factor of hundred by allowing the neutron beam to pass through a slit made of the $^6$Li tiles, which has a thickness of 9.6 mm and a pinhole with a diameter of 1.5 mm.

\begin{figure}[ht]
\centering
  \includegraphics[width=8cm]{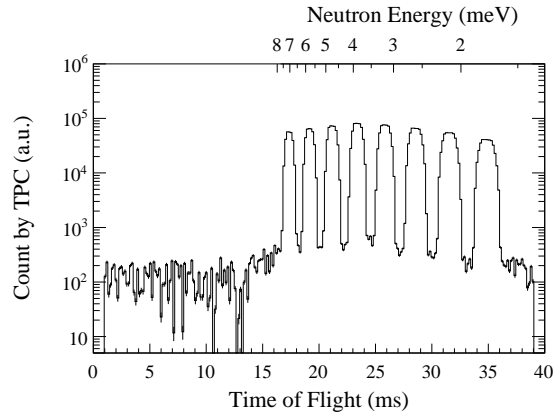} 
 \caption{Time of flight spectrum of the bunched neutron beams measured using the TPC.
 }
  \label{fig_TOFplot}
\end{figure}

In Fig.~\ref{fig_TOFplot}, the time of flight distribution of the reaction events for the bunched neutron beams, as counted by the TPC, is shown.
The origin, in Fig.~\ref{fig_TOFplot}, is the time at which pulsed neutrons were generated at the mercury target.
The energy width of the pulsed neutron beams was estimated to be 1--7 meV using the time of flight information and the distance between the TPC and the mercury target as 20 m.
Eight bunches of neutrons were generated by the SFC in each neutron pulse.
The interval length of these bunches was determined as 1.7 m, so that a new bunch was injected into the TPC only after the previous bunch reached the beam catcher, for counting the reaction events occurring in each bunch.
The average flux of the incident neutron beam was found to be $2.1\times10^3$ neutrons/s at the proton beam power of 170 kW in this experiment.
The duty factor of the neutron bunch was calculated as 0.095. 

The measurements taken with the neutron injection for a duration of 1500 s (``open'' data), and without the neutrons for a duration of 150 s (``closed'' data), were repeated alternately to subtract the beam-independent background.
The actual measurement times for the gases I and I\hspace{-.1em}I were 24 hours and 75 hours, respectively.
\clearpage
\subsection{DETECTOR}
\begin{figure}[ht]
\centering
  \includegraphics[width=10cm]{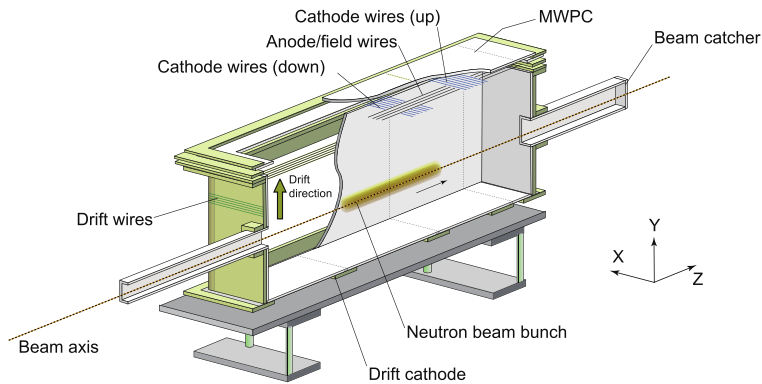} 
  \caption{Schematic view of the TPC~\cite{Arimoto2015}. 
}
  \label{fig_tpc}
\end{figure}

\begin{figure}[ht]
\centering
  \includegraphics[width=8cm]{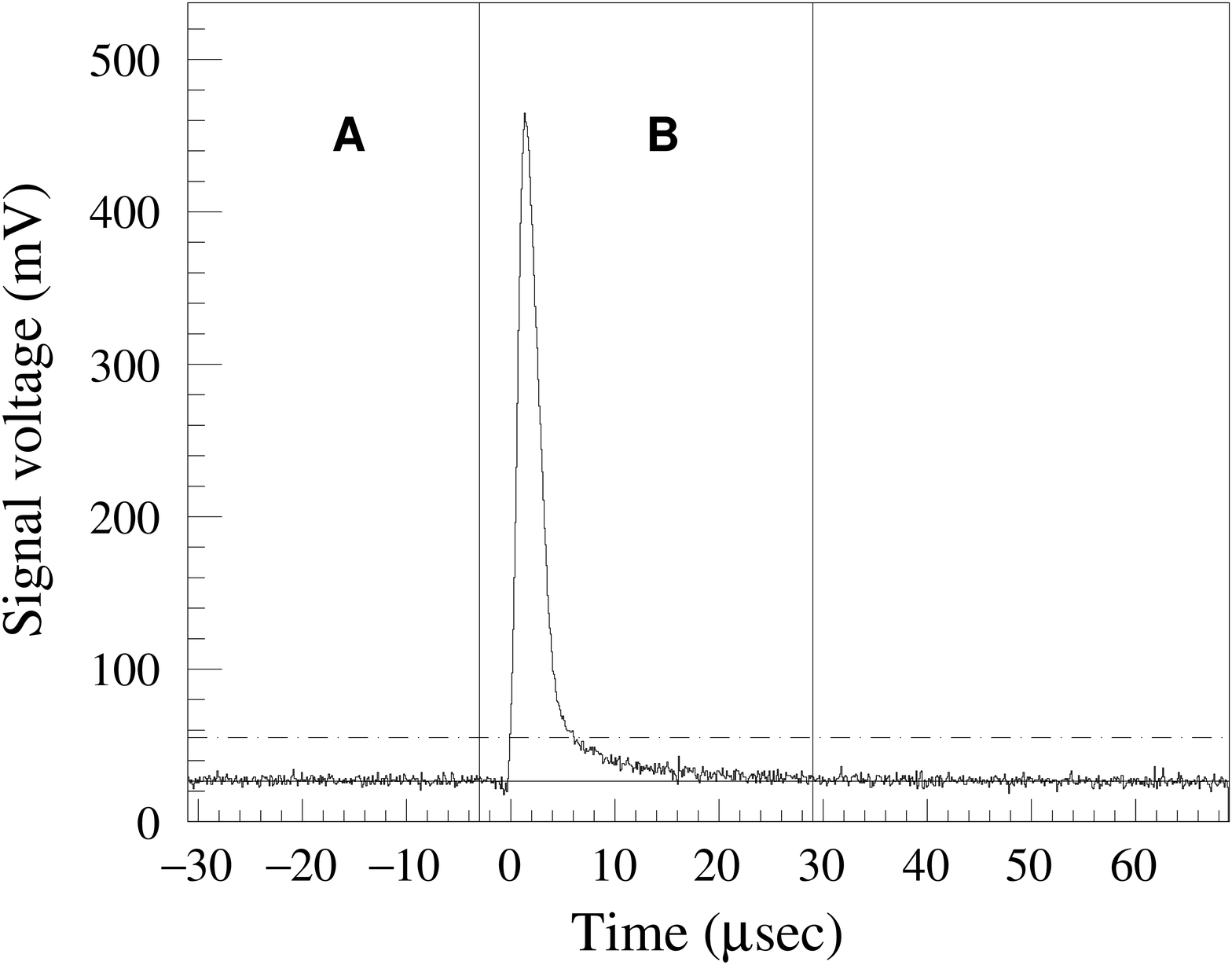} 
  \caption{
  Typical shape of a triggered anode signal as a function of time. The time at which the event is triggered is chosen as the origin of the time axis. In region A without the envelope of signal peak, the baseline (horizontal solid line at $\sim$20 mV) was determined by taking an average of the signal voltage. In region B, where the pulse height exceeds the baseline voltage of 30 mV (dash-dotted line), the wire channel is treated as a hit channel.
  }
  \label{fig_waveform}
\end{figure}

\begin{figure}[ht]
\centering
  \includegraphics[width=8cm]{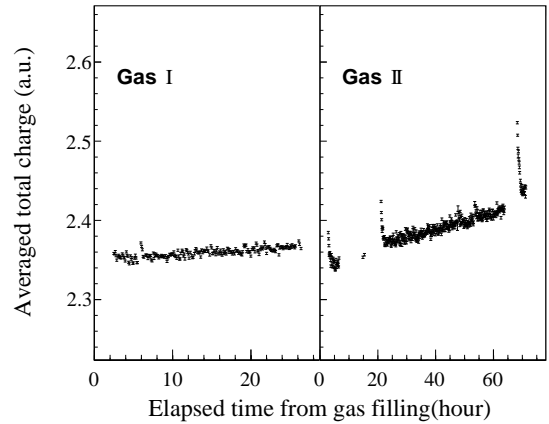} 
  \caption{Time dependences of the peak charges for  $^{14}$N(n,p)$^{14}$C events in gas I (left) and in gas I\hspace{-.1em}I (right).}
  \label{fig_asumtime}
\end{figure}

The TPC shown in Fig.~\ref{fig_tpc} was originally developed for the neutron lifetime measurement~\cite{Arimoto2015}.
Thus, some parameters were optimized for this measurement.
The size of the sensitive region in the TPC was selected as 290 mm ($X$) $\times$ 300 mm ($Y$) $\times$ 960 mm ($Z$)~\cite{Arimoto2015}.
To avoid any electrical discharges in the TPC, a uniform electric field of 300 V/cm was applied in the drift volume, that produced the same voltages as used in the set-up for the neutron lifetime measurement.
The multiwire proportional chamber was attached at the top of the drift volume to detect the electrons generated along tracks and was constructed with three layers of sensitive wires~\cite{Arimoto2015}.
The anode wires and the field wires were attached alternately at every 6 mm-pitch along the $Z$-axis in the central layer.
Cathode wires were attached at every 6 mm-pitch along the $X$-axis in the upper and lower layers.
A voltage of 1520 V was applied to the anode wires and 0 V was applied to the other wires. 
The gain of the TPC was calibrated using the peak position of the $^{14}$N(n,p)$^{14}$C reaction in the total charge spectra of the ``open'' data.
The pulse shape of each wire signal every event was recorded as shown in Fig.~\ref{fig_waveform}.
The total charge of an individual event is calculated by integrating the pulse height above the baseline relative to the leading edge at the pulse height of approximately 30 mV (as seen in Fig.~\ref{fig_waveform}), in the time interval of -3 ${\rm \mu s}$ to 29 ${\rm \mu s}$.
The gain drift was monitored at a regular interval of 300 s.
Figure~\ref{fig_asumtime} shows the gain drift of the average total charge for the reaction events detected within each time period of monitoring the events.
The total charge for each event was corrected using the interpolated gain drift.

\begin{figure}[hh]
\centering
  \includegraphics[width=15cm]{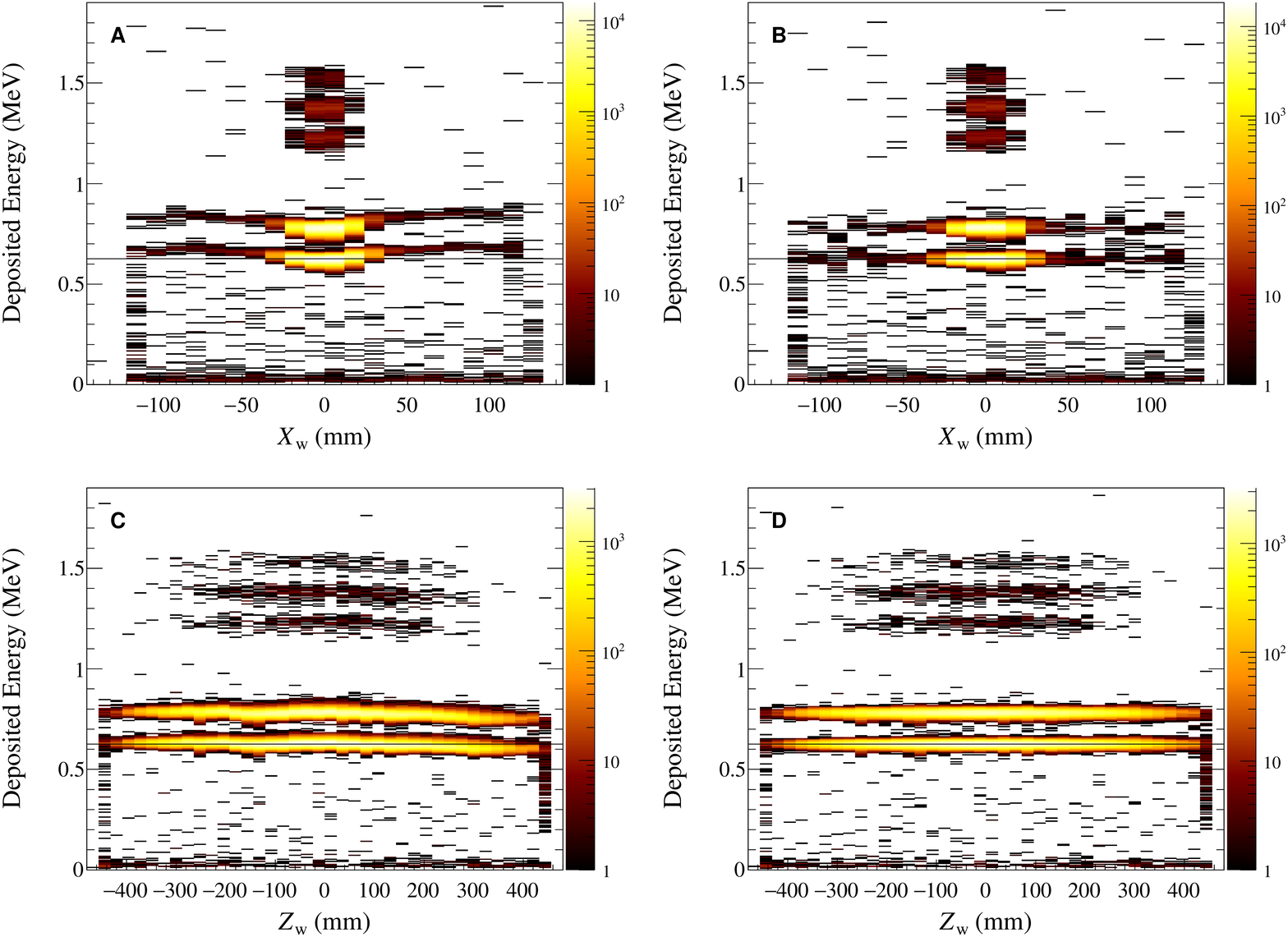} 
  \caption{
  $X_{\mathrm{w}}$ and $Z_{\mathrm{w}}$ dependence of the deposited energy in gas I shows;(A) $X_{\mathrm{w}}$ before the compensation, (B) $X_{\mathrm{w}}$ after the compensation, (C) $Z_{\mathrm{w}}$ before the compensation, and (D) $Z_{\mathrm{w}}$ after the compensation. The origin for each corresponds to the center of the TPC. The horizontal lines indicate the $^{14}$N(n,p)$^{14}$C reaction energy as 0.626 MeV. (A) shows a decrease in the peak height at the center and near the edges of the sensitive volume, which is attributed to the gain saturation due to the high count rates in the regions located on the beam axis and near the inner walls of the TPC.
  }
  \label{fig_poscalib}
\end{figure}

In order to compensate for the dependence of the deposited energy on the $X$ and $Z$ positions, the energy centers of the tracks, $X_{\mathrm{w}}$ and $Z_{\mathrm{w}}$, were evaluated using the anode wires and the high gain cathode wires from the lower layer in Fig.~\ref{fig_tpc}.
Figure~\ref{fig_poscalib} shows the $X_{\mathrm{w}}$ and $Z_{\mathrm{w}}$ dependences of the deposited energy before and after the compensation.
After the compensation for the position dependence of the gain, a uniform spatial distribution of the TPC gain was obtained as shown in Fig.~\ref{fig_poscalib} B and Fig.~\ref{fig_poscalib} D.

\clearpage
\subsection{GAS HANDLING SYSTEM}
\label{gashandlingsystem}
\begin{figure}[ht]
\centering
  \includegraphics[width=12cm]{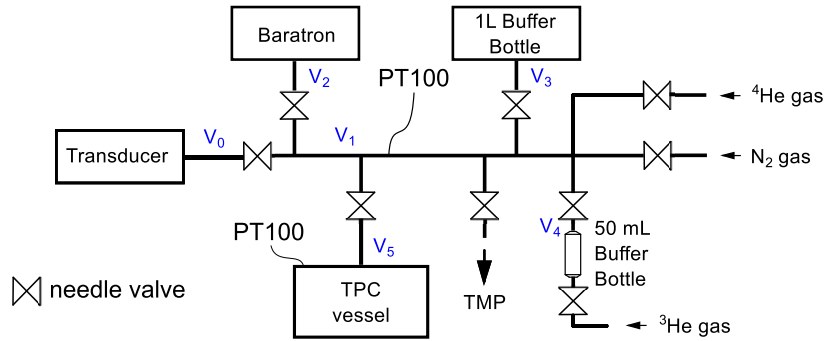} 
  \caption{Schematic view of the gas handling system~\cite{HESJ_G3}. 
 }
  \label{fig_gasline}
\end{figure}

The gas injection and the measurement of $\rho(^{3} \mathrm{He})/\rho(^{14} \mathrm{N})$ were performed using the gas handling system~\cite{HESJ_G3}, essentially the same system that is also used for the neutron lifetime measurement.
Figure~\ref{fig_gasline} shows a schematic view of the gas handling system used in this work.
The gas handling system consists of five volumes ($V_0$-$V_4$): $V_0$, $V_1$, and $V_2$ are made up of stainless-steel tubes; $V_3$ is a 1-L buffer bottle used for measuring the volume ratio; $V_4$ is a 50-mL storage bottle, in which the isopure $^3$He gas (greater than 99.95\% purity, provided by ISOTEC) was stored; $V_5$ is the volume of the TPC vessel.
The values of $V_0$-$V_5$ are shown in Table~\ref{tab_volume}, as measured by a manometer.
The volume, $V_5$ was evaluated using the results of $V_0$-$V_4$ measurements combined with the volume ratio measurement as shown in Table~\ref{tab_vratio}. 
The cylinders of G1He (impurity 5 ppm) and natural N$_2$ (impurity 1.9 ppm) gases, provided by Tomoe Shokai, were connected to $V_1$.
The impurity of the N$_2$ gas slightly increased with the contamination from the atmosphere due to possible leakage at the nylon tube that is used for connecting the N$_2$ gas cylinder to the gas handling system.
The $^3$He content ratio in the G1He gas was measured as 0.111(2) ppm by a mass spectrometer~\cite{Sumino2001}.
In this mass spectrometer, the synthesized helium gas with a $^3$He content ratio of 27.36(11) ppm~\cite{HESJ_G3} was used as a primary standard to correct for the $^3$He/$^4$He discrimination effect.
A piezoresistive transducer and a Baratron gauge~\cite{HESJ_G3} were included in $V_0$ and $V_2$, respectively.
The temperature of the gas handling system was monitored using a platinum resistance thermometer sensor (PT100) attached to $V_1$.
Another PT100 thermometer was placed on the TPC vessel to monitor the representative temperature of the gas in the TPC vessel.
\begin{table}[htp]
\caption{Volumes of gas handling system}
\label{tab_volume}
\begin{center}
\begin{tabular}{c c}\hline
Parameter & Volume (cm$^3$) \\\hline\hline
$V_0$ & 43.0(3)  \\
$V_1$ & 95.6(3)  \\
$V_2$ & 14.6(1) \\
$V_3$ & 1004(3)\\
$V_4$ & 57.0(3)\\
$V_5$ & 6.37(4)$\times 10^{5}$\\\hline
\end{tabular}
\end{center}
\end{table}

To estimate $\rho(^{3} \mathrm{He})/\rho(^{14} \mathrm{N})$ for the cross section measurement, we regarded the Boltzmann constant as unity in this work.
The uncertainty in the pressure ($P$) to the temperature ($T$) ratio is due to the measurement uncertainties associated with the thermometers and the pressure gauges~\cite{HESJ_G3} used in the experiment.
The measured values of $P/T$ and the uncertainties associated with these measurements are shown in Table~\ref{tab_numratio}.
The low $^{3}$He pressure of approximately 10 Pa was determined using two independent approaches, for reliability. The first approach was a direct measurement (DM) using the Baratron gauge, and the second one involved the volume expansion (VE) method.
The principle of the VE method is that a low pressure of gas after diluting in a larger volume is evaluated by the high pressure before diluting in a smaller volume and the volume ratio.
The procedure for the measurement of the number densities is as follows.
The high pressure of the $^3$He gas was measured in a small volume of $V_{\rm {ini}}=V_0+V_1+V_4$.
Then, the $^3$He gas was released to a larger volume of $V_{\rm {fin}}=V_0+V_1+V_2+V_4+V_5$.
The initial number density of $\rho_{\rm ini}\propto 1/V_{\rm {ini}}$ is calculated by the state equation and the final number density of $\rho_{\rm fin}\propto 1/V_{\rm {fin}}$ is obtained using the relation: $\rho_{\rm fin}=\rho_{\rm ini}R_{\rm V}$, where $R_{\rm V}=V_{\rm {ini}}/V_{\rm {fin}}$ is the volume ratio.
Three types of volume ratios, $R_{1-3}$ were measured with the gas handling system as shown in Table~\ref{tab_vratio}, to determine the value of $R_\mathrm{V}$ using Eq.~\ref{eqRv}, as shown below. 
\begin{eqnarray}
\label{eqRv}
R_\mathrm{V}=\frac{V_0+V_1+V_4}{V_0+V_1+V_2+V_4+V_5}=\left[\frac{R_2}{R_1 R_3} - \frac{R_2}{R_1} + 1 \right]^{-1}.
\end{eqnarray}
A more detailed description of this method can be found in Ref.~\cite{HESJ_G3}.
When the temperatures of the gas handling system and the TPC vessel in the volume ratio measurement are different from those in the cross section measurement, the volume ratio is corrected with the thermal expansion condition of the volumes.
Then, the effective volume ratio, $R^{\prime}$ is expressed as,
\begin{eqnarray}
\label{eqRvprime}
R_\mathrm{V}^{\prime}=R_V \frac{T_1}{T_2}\frac{T^{\prime}_2}{T^{\prime}_1},
\end{eqnarray}
where $T_1$ and $T_2$ are the temperatures at $V_1$ and $V_5$, respectively, in the volume ratio measurement; $T^{\prime}_1$ and $T^{\prime}_2$ are those in the cross section measurement.
\begin{table}[htp]
\caption{Results of volume ratio measurement.}
\label{tab_vratio}
\begin{center}
\begin{tabular}{c c c}\hline
Parameter & Volumes & Ratio \\\hline\hline
$R_1$ & ($V_0$+$V_1$)/($V_0$+$V_1$+$V_3$) & 0.12032(4) \\
$R_2$ & ($V_0$+$V_1$)/($V_0$+$V_1$+$V_4$) & 0.70558(14)\\
$R_3$ & ($V_0$+$V_1$+$V_3$)/($V_0$+$V_1$+$V_2$+$V_3$+$V_5$) & $1.8081(24) \times 10^{-3}$\\\hline\hline
$R_\mathrm{V}$ & ($V_0$+$V_1$+$V_4$)/($V_0$+$V_1$+$V_2$+$V_4$+$V_5$) & $3.088(6)\times10^{-4}$\\\hline
\end{tabular}
\end{center}
\end{table}

In the VE method, the pressure of the gas handling system, $P_\mathrm{VE}\sim$30 kPa was measured, when the $^{3}$He gas was released to $V_{\rm ini}$.
The number density of the $^{3}$He nuclei, $\rho_{\mathrm{VE}}(^3 \mathrm{He})$ was calculated using the state equation (Eq.~\ref{eqVE}),
\begin{eqnarray}
\label{eqVE}
\rho(^{3} \mathrm{He})_{\mathrm{VE}}=\frac{P_{\mathrm{VE}}R^{\prime}_\mathrm{V}}{Z_{^3\mathrm{He}}T_{\mathrm{VE}}},
\end{eqnarray}
where $T_\mathrm{VE}$ is a representative gas temperature measured at $V_1$ ($T^{\prime}_1=T_\mathrm{VE}$); $Z_{^3\mathrm{He}}$ is the compressibility factor to compensate for the discrepancy from the state equation of the ideal gas, and it was calculated using the second virial coefficient of helium gas as 11.83(3) cm$^3$/mole \cite{Kell1978}.
After $^{3}$He was released to $V_{\rm {fin}}$, the pressure of the gas handling system, $P_\mathrm{DM}\sim$10 Pa was measured using the Baratron gauge.
The number density of the $^{3}$He nuclei, $\rho_{\mathrm{DM}}(^3 \mathrm{He})$ was determined, using the pressure value from the DM, in the following state equation (Eq.~\ref{eqDM}),
\begin{eqnarray}
\label{eqDM}
\rho_{\mathrm{DM}}(^3 \mathrm{He})=\frac{P_{\mathrm{DM}}C_{\mathrm{th}}}{ T_{\mathrm{DM}}},
\end{eqnarray}
where $T_\mathrm{DM}$ is the representative gas temperature measured at $V_5$ ($T^{\prime}_2=T_\mathrm{DM}$); $C_{\mathrm{th}}$ is the correction factor of the thermal transpiration effect for helium gas~\cite{Setina1999}.
The measured value of pressure (DM) from the Baratron gauge is different from the pressure inside the TPC vessel due to the thermal transpiration effect.
The correction factor, $C_{\mathrm{th}}$ depends on the gas species and the ratio of the mean free path, $\lambda$ of measured gas to a diameter, $d$ of the tube connecting the Baratron to the gas handling system.
In the molecular regime ($\lambda/d \gg 1$), the correction factor is the square root of ratio of temperatures of the Baratron gauge (at 318 K) and the TPC (at $\sim$300 K), $\sqrt{300/318}\sim0.97$, whereas in the viscous regime ($\lambda/d \ll 1$)the fa,  is alctormost unity.
In this measurement, $C_{\mathrm{th}}$ was calculated as 0.98--0.99 because the experimental condition was intermediate between the molecular and viscous regimes.

First the N$_2$ gas, and then the G1He gas were injected after $^3$He gas introduction, and a piezoresistive transducer was used to measure their filling pressures, $P_{\mathrm{G1He}}$ and $P_{\mathrm{N}_2}$, respectively.
The typical gas temperatures, $T_{\mathrm{G1He}}$ and $T_{\mathrm{N}_2}$ were measured at $V_5$ after the injection of each gas.
$\rho(^{14} \mathrm{N})$ was determined using the following state equation (Eq.~\ref{eqN});
\begin{eqnarray}
\label{eqN}
\rho(^{14} \mathrm{N})=\frac{2 P_{\mathrm{N}_2}A_{^{14}\mathrm{N}}}{Z_{\mathrm{N}}T_{\mathrm{N}_2}C_{\mathrm{def}}},
\end{eqnarray}
where $Z_{N}$ is the compressibility factor of N$_2$ gas as calculated using the second virial coefficient of nitrogen gas as -4.2(5) cm$^3$/mole \cite{Sevast1986}; $A_{^{14}\mathrm{N}}$ is the isotope ratio of 0.9964(2) in the natural N$_2$ gas as recommended by IUPAC~\cite{Berglund2011}; $C_{\mathrm{def}}$ is the correction factor for the deformation of the TPC vessel, as the volume of the vessel changed before and after the injection of N$_{2}$ and G1He gases.
To prepare gas I, the G1He gas was injected first, whereas the order of the gas injections was reversed for gas I\hspace{-.1em}I.
The total pressure of gas I was 100 kPa and that of gas I\hspace{-.1em}I was 20 kPa, at the end of N$_2$ gas injection.
The degree of the deformation of the TPC vessel was estimated from the total pressure and calculation of the upper limit for the strain displacement of the TPC vessel.
Thus, taking the calculated value as the upper limit for the correction, the target volume was corrected for half the maximum deformation, with the same amount assigned as the uncertainty of the correction.
The correction factors $C_{\mathrm{def}}$ and their uncertainty for deformation in gas I and gas I\hspace{-.1em}I were calculated as 0.9985(15) and 0.9997(3), respectively.
In order to minimize the corrections in future, we suggest that N$_{2}$ gas should be injected prior to G1He gas. 
The value of $\rho(^{3} \mathrm{He})_{\mathrm{G1He}}$ was calculated as $2.96(5) \times 10^{-5}$ Pa/K, using the $^3$He content ratio, $P_{\mathrm{G1He}}$, $T_{\mathrm{G1He}}$, and the compressibility factor of $Z_{\mathrm{G1He}}$.

\begin{figure}[ht]
\centering
  \includegraphics[width=8cm]{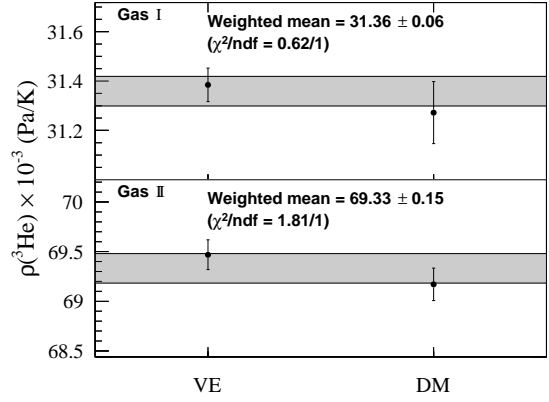} 
  \caption{
    Number density of the $^3$He nuclei (extracted from the $^3$He cylinder) calculated as the weighted mean of the values obtained from both the methods; the volume expansion (VE) and the direct measurement (DM). The gray band shows the weighted mean and the scaled error.
    }
  \label{fig_hecombine}
\end{figure}

The calculated values of $\rho(^{3} \mathrm{He})_{\mathrm{VE}}$ and $\rho(^{3} \mathrm{He})_{\mathrm{DM}}$ have the same order-of-magnitude accuracy.
Hence, the number density of the $^3$He nuclei from the isopure $^3$He gas, $\rho(^{3} \mathrm{He})_{\mathrm{WM}}$ was calculated as a weighted mean of the values obtained from both the methods (DM and VE) as shown in Fig.~\ref{fig_hecombine}.
The value of the number density possesses a scaled error~\cite{HESJ_G3,Olive2016}. 
Finally, the total number density of the $^3$He nuclei in Eq.~\ref{eqNsigma} was calculated as $\rho(^{3} \mathrm{He})=\rho_{\mathrm{WM}}(^{3}\mathrm{He}) + \rho(^{3}\mathrm{He})_{\mathrm{G1He}}$.
$\rho(^{3} \mathrm{He})/\rho(^{14} \mathrm{N})$ with an uncertainty of 0.3\% as shown in Table~\ref{tab_numratio}.

\begin{table}[htp]
\caption{Results of the number density ratio $\rho({^{3} \mathrm{He}})/\rho({^{14} \mathrm{N}})$ evaluation.}
\label{tab_numratio}
\begin{center}
\begin{tabular}{c c c}\hline
 & \multicolumn{2}{c}{ Value and Uncertainty}  \\
Factor & Gas I & Gas I\hspace{-.1em}I \\\hline\hline
$P_{\mathrm{DM}}/T_{\mathrm{DM}}$ (Pa/K) & 0.03182(11) & 0.06992(14) \\
$C_{\mathrm{th}}$ & 0.9826(17) & 0.9892(11) \\
$\rho(^{3} \mathrm{He})_{\mathrm{DM}}$ (Pa/K) & 0.03127(13) & 0.06917(16) \\\hline
$P_{\mathrm{VE}}/T_{\mathrm{VE}}$ (Pa/K) & 100.99(3) & 223.42(7) \\
{\bf $R_{\mathrm V}^{\prime}$ }& $3.108(7)\times10^{-4}$ & $3.110(7)\times10^{-4}$ \\
$Z_{^3\mathrm{He}}$ & 1.0001437(4) & 1.0003179(8) \\
$\rho(^{3} \mathrm{He})_{\mathrm{VE}}$ (Pa/K) & 0.03138(7) & 0.06947(15) \\\hline
the purity of $^{3}$He (\%) & $^{+0.00}_{-0.05}$ & $^{+0.00}_{-0.05}$ \\
$\rho(^{3} \mathrm{He})_{\mathrm{WM}}$ (Pa/K) & 0.03136(6) & 0.06933(15) \\\hline
$P_{\mathrm{G1He}}/T_{\mathrm{G1He}}$ (Pa/K) & 267.30(13) & 266.91(13) \\
$Z_{\mathrm{G1He}}$ & 1.0003803(10) & 1.0003798(10) \\
$\rho(^{3}\mathrm{He})_{\mathrm{G1He}}$ (Pa/K) & $2.96(5) \times 10^{-5}$ & $2.96(5) \times 10^{-5}$ \\\hline\hline
$\rho(^{3} \mathrm{He})$ (Pa/K) & 0.03139(6) & 0.06936(15) \\\hline\hline
$P_{\mathrm{N}_2}/T_{\mathrm{N}_2}$ (Pa/K) & 66.90(8) & 66.76(5) \\
$Z_\mathrm{N}$ & 0.999966(4) & 0.999966(4) \\
the purity of N$_2$ (ppm) & $^{+0.0}_{-1.9}$ & $^{+0.0}_{-1.9}$ \\
$C_{\mathrm{def}}$ & 0.9985(15) & 0.9997(3) \\\hline\hline
$\rho(^{14} \mathrm{N})$ (Pa/K) & 133.5(3) & 133.07(11) \\\hline\hline
$\rho(^3 \mathrm{He})/\rho(^{14} \mathrm{N})$ & $2.351(7) \times 10^{-4}$ & $5.212(12) \times 10^{-4}$ \\\hline
\end{tabular}
\end{center}
\end{table}
\clearpage
\section{ANALYSIS}
\label{ANALYSIS}
\subsection{EVENT SELECTION}
\label{eventselection}
\begin{figure}[p]
\centering
  \includegraphics[width=12cm]{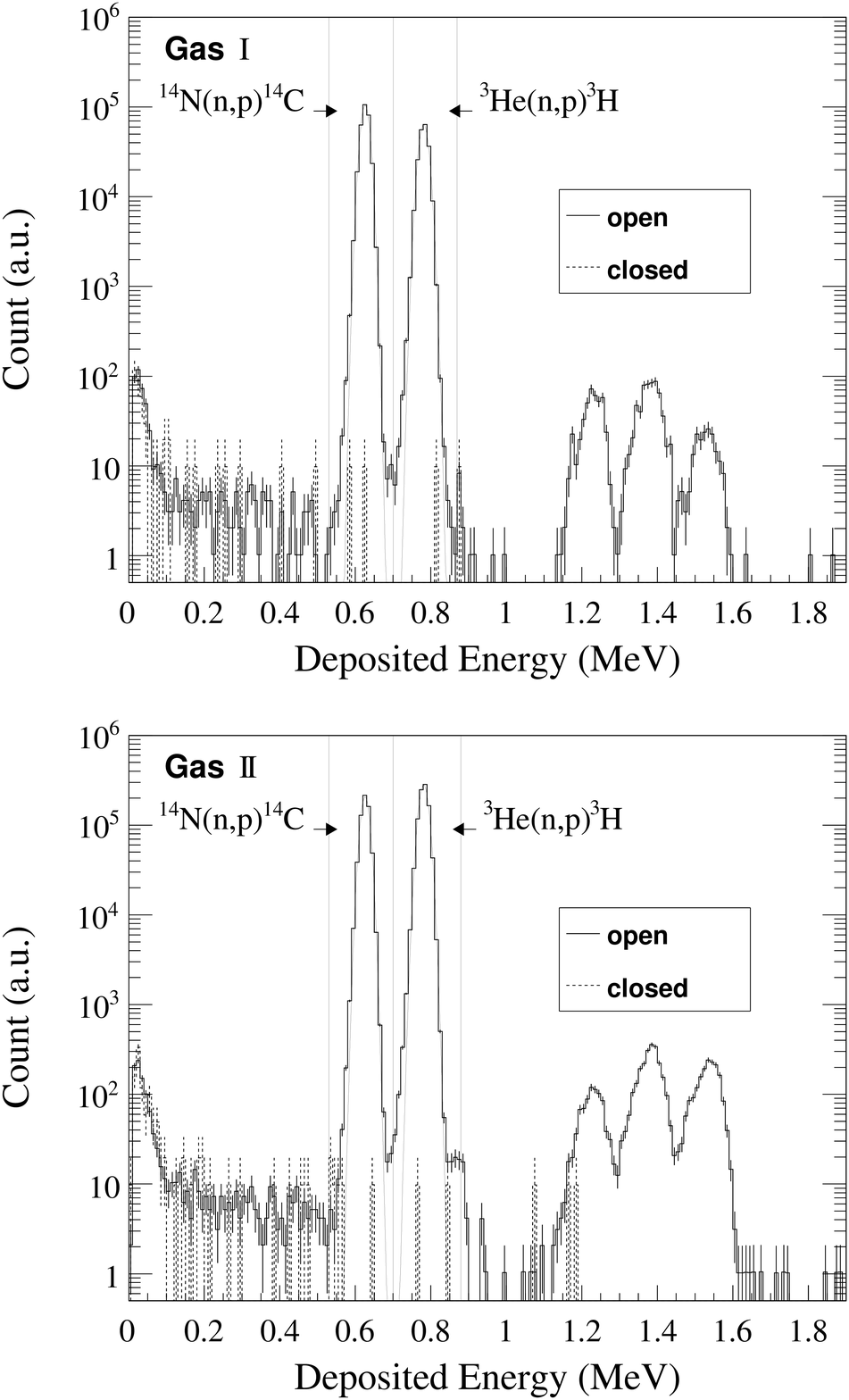} 
  \caption{
    Measured deposited energy distributions for gas I (top) and gas I\hspace{-.1em}I (bottom). Solid lines indicate the results with the ``open'' data, and the dotted lines indicate the results with the ``closed'' data. The number of the reaction events in the ``closed'' data were scaled by the neutron flux measured with the beam monitor.
  }
  \label{fig_depE}
\end{figure}

In this work, the events corresponding to the $^{14}$N(n,p)$^{14}$C and $^{3}$He(n,p)$^{3}$H reactions were identified with their deposited energies in the TPC. 
The candidate events of both the reactions were selected by the following conditions.
Firstly, the events corresponding to both the reactions, that produced their complete tracks within the TPC, were extracted by setting the window dimensions as $\vert$$X_{\mathrm{w}}$$\vert \leqq$ 72 mm and -432 mm $\leqq$ $Z_{\mathrm{w}}$ $\leqq$ 408 mm.
Secondly, the events whose falling time of the signal was less than 29 $\mathrm{\mu}$s (the end of the region B as shown in Fig.~\ref{fig_waveform}) were selected to minimize the double-counted events.
Finally, the events in the TOF gate when the bunched neutron beams (as shown in Fig.~\ref{fig_TOFplot}) were present completely inside the sensitive volume of the TPC (-432 mm $\leqq$ $Z_{\mathrm{w}}$ $\leqq$ 432 mm) were selected to reduce the background events caused by the neutron capture reactions occurring around the TPC.
The distributions of the deposited energy observed using the above-mentioned conditions are shown in Fig.~\ref{fig_depE} for gas I (top) and gas I\hspace{-.1em}I (bottom).
The double gaussian peak fitting of this data yielded the relative standard deviation ($\sigma/{\rm mean}$) of the $^{14}$N(n,p)$^{14}$C and the $^{3}$He(n,p)$^{3}$H reactions as 1.8\% and 1.6\%, respectively.

The values of $S(^{14} \mathrm{N})$ and $S(^3 \mathrm{He})$ were obtained from the number of events with their energy peaks around 0.626 MeV and 0.764 MeV, respectively (as shown in Fig.~\ref{fig_depE}).
The internal boundary of the energy peak profile, between two peaks, was determined as the local minimum point, approximately 7$\sigma$ away from the center of each peak.
The external boundary was determined as 8$\sigma$ away from the center of each peak.
These boundary conditions allow for the entire energy peak to be contained within each integrated range.
A few small but significant events were observed around the local minimum point ($\sim$0.7 MeV). The source of such events is suspected due to the reaction products escaping the TPC sensitive volume, sharing the total reaction energy to a small signal peak below the threshold, or double-counting with a low deposited energy event (i.e. gamma-ray event), however it could not be identified.
Thus, the maximum numbers of unclassified events are estimated by multiplying the average event per energy in 0.69--0.71 MeV and the energy range of integration for each peak, and regarded as the systematic uncertainties in the numbers of events.
The systematic uncertainty of $S(^{14} \mathrm{N})/S(^3 \mathrm{He})$ was found to be $\pm0.1$\%.
Table~\ref{tab_eratio} summarizes the value of each parameter and the uncertainty associated with each.

The events in the ``closed" data are mainly caused by the cosmic muons and the low energy peak at $\sim$0.1 MeV appeared due to the Compton scattering events caused by the prompt gamma-rays from the upper stream.
Since, the total number of the background events in the energy peaks contributed less than 0.01\% uncertainty to the values of $S(^{14} \mathrm{N})$ and $S(^3 \mathrm{He})$, as evaluated by the ``closed" data, therefore such background events were ignored.
As seen in Fig.~\ref{fig_depE}, we observed three peaks in the vicinity of ~1.4 MeV due to composite double-counted events from the $^{14}$N(n,p)$^{14}$C and $^{3}$He(n,p)$^{3}$H reactions. 
However, the values of $S(^{14} \mathrm{N})$ and $S(^3 \mathrm{He})$ were calculated using the number of the single-counted events only and the double-counted events in the integrated range were rejected by the extraction condition of the falling time of the signal.
\begin{table}[htp]
\caption{Reaction event ratio $S(^{14} \mathrm{N})/S(^{3} \mathrm{He})$.}
\label{tab_eratio}
\begin{center}
\begin{tabular}{c c c}\hline
& \multicolumn{2}{c}{Value and Uncertainty} \\
Effect & Gas I & Gas I\hspace{-.1em}I \\\hline\hline
$S(^{14}\mathrm{N})$ (count)& $291825 \pm 540$ & $585745 \pm 765 $\\
$S(^{3}\mathrm{He})$ (count)& $195655 \pm 442$ & $872631 \pm 934$ \\\hline
unclassified events & &\\
$^{14}$N(n,p)$^{14}$C (count)&105 & 363\\
$^{3}$He(n,p)$^{3}$H (count) & 105 & 364 \\\hline\hline
&  1.4915 & 0.6712 \\ 
$S(^{14} \mathrm{N})/S(^3 \mathrm{He})$  & $\pm 0.0044$ (stat.) & $\pm 0.0011$ (stat.) \\ 
& $\pm 0.0013$ (sys.)  & $\pm 0.0007$ (sys.) \\ \hline
\end{tabular}
\end{center}
\end{table}
\clearpage
\subsection{DITECTION EFFICIENCY}
The efficiency ratio, $\varepsilon(^{3}\mathrm{He})/\varepsilon(^{14} \mathrm{N})$ was evaluated from the experimental data.
As shown in Fig.~\ref{fig_waveform}, the typical pulse heights for the $^{14}$N(n,p)$^{14}$C and $^{3}$He(n,p)$^{3}$H events were 400 mV above the threshold values, therefore the trigger efficiencies were regarded as approximately 100\% for each case.
Additionally, the difference in the efficiencies, obtained from the extraction conditions for both the reactions, is insignificant.
Since both the reactions were measured in the TPC, the extraction efficiency of the TOF gate and the falling time of the signal should be the same for the two reactions.

\begin{figure}[th]
\centering
  \includegraphics[width=15cm]{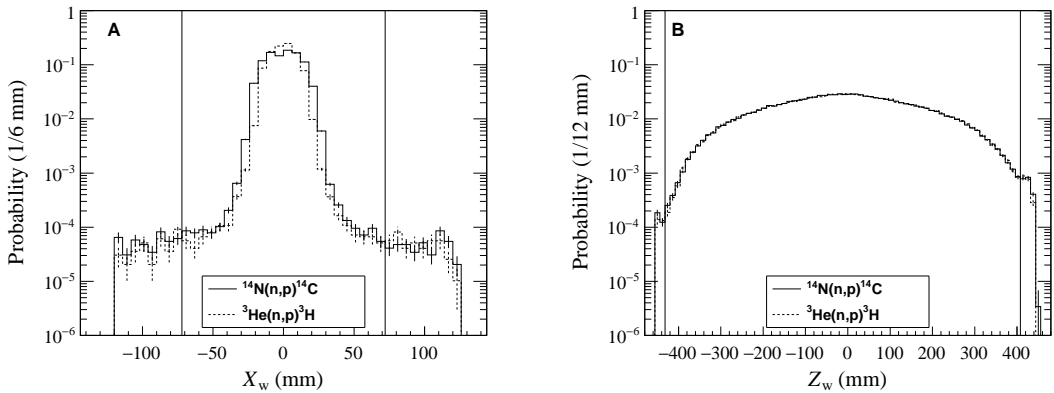} 
  \caption{
    Center of deposited energy for the reactions in gas I with normalized distribution  of (A) $X_{\mathrm{w}}$ and (B) $Z_{\mathrm{w}}$. Solid lines indicate the total events of the $^{14}$N(n,p)$^{14}$C reaction, and dotted lines indicate those of the $^{3}$He(n,p)$^{3}$H reaction. Vertical lines indicate the boundaries of the extraction condition used for the event selection.
  }
  \label{fig_ce}
\end{figure}

The difference in the extraction efficiencies between both the events was derived using the extraction efficiencies of $X_{\mathrm{w}}$ and $Z_{\mathrm{w}}$ because the track lengths of the reaction products and the deposited energies for the two reactions were different.
The calculated distances (by \verb|SRIM|~\cite{SRIM}) from the reaction point to the weighted center of the deposited energies for the $^{14}$N(n,p)$^{14}$C and the $^{3}$He(n,p)$^{3}$H reactions were 16.3 mm and 11.4 mm, respectively.
In Fig.~\ref{fig_ce}, the normalized distributions; $X_{\mathrm{w}}$ and $Z_{\mathrm{w}}$, with respect to the integrated peak counts of the $^{14}$N(n,p)$^{14}$C and the $^{3}$He(n,p)$^{3}$H events, respectively, are presented.
As shown in Fig.~\ref{fig_ce} A, the difference between the root mean square $X_{\mathrm{w}}$ values for the $^{14}$N(n,p)$^{14}$C and $^{3}$He(n,p)$^{3}$H reactions is 2.4 mm, which corresponds to the result obtained from the \verb|SRIM| calculations with isotropic emission.
Since the number of the $^{14}$N(n,p)$^{14}$C reaction events occurring outside the extraction region was greater, the effective efficiency for the $^{14}$N(n,p)$^{14}$C reaction was lower than that for the $^{3}$He(n,p)$^{3}$H reaction.
The upper limit for the corrected value of the effective efficiency ratio was estimated by the event probabilities in the 6-mm bin widths on either side of the boundaries of the extraction conditions as shown in Fig.~\ref{fig_ce} A and B. 
The corrected value was regarded also to represent the uncertainty in the efficiency ratio $\varepsilon(^{3}\mathrm{He})/\varepsilon(^{14} \mathrm{N})$.
The total corrected value of $\varepsilon(^{3}\mathrm{He})/\varepsilon(^{14} \mathrm{N})$ was evaluated to be 0.07\%, and it was regarded also to be the uncertainty of $\varepsilon(^{3}\mathrm{He})/\varepsilon(^{14} \mathrm{N})$.
These values are shown in Table~\ref{tab_eff}.
\begin{table}[htp]
\caption{Extraction efficiency ratio $\epsilon_{^{3} \mathrm{He}}/\epsilon_{^{14} \mathrm{N}}$.}
\label{tab_eff}
\begin{center}
\begin{tabular}{c c c}\hline
& \multicolumn{2}{c}{Value $\pm$ Uncertainty} \\
Effect & Gas I & Gas I\hspace{-.1em}I \\\hline\hline
Escape along $X$-direction & $(1.8 \pm 1.8) \times 10^{-4}$ & $(1.4 \pm 1.4) \times 10^{-4}$ \\
Escape along $Z$-direction & $(7.0 \pm 7.0) \times 10^{-4}$ & $(6.3 \pm 6.3) \times 10^{-4}$ \\\hline\hline
$\varepsilon(^{3} \mathrm{He})/\varepsilon(^{14} \mathrm{N})$& $1.0007 \pm 0.0007 $& $1.0007 \pm 0.0007$ \\\hline
\end{tabular}
\end{center}
\end{table}
\section{RESULT}
\begin{figure}[th]
\centering
  \includegraphics[width=15cm]{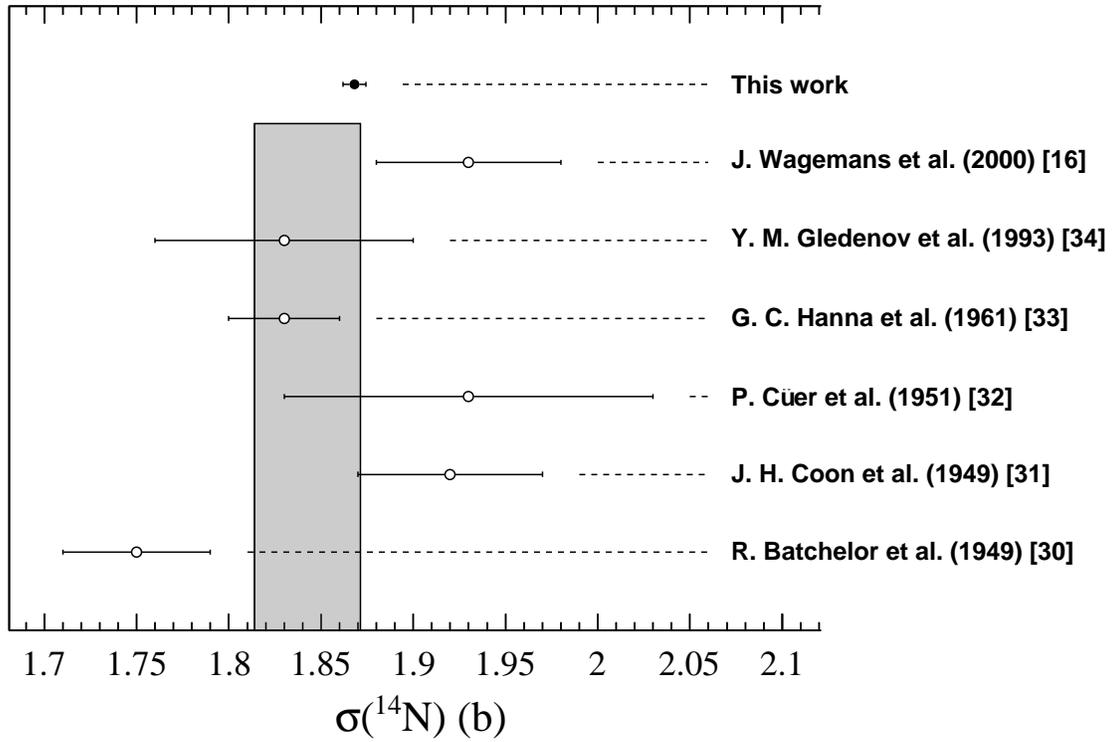} 
  \caption{
    Experimental results of $\sigma({^{14}\mathrm{N}})$, the cross section of the $^{14}$N(n,p)$^{14}$C reaction at the neutron velocity of 2200 m/s. The weighted mean from the previous results (open circles) \cite{Batch1949,Coon1949,Cuer1951,Hanna1961,Gledenov1993,14NWagemans2000} was $1.84 \pm 0.03$ b with $\chi^2/ndf$ of 11.78/5 (gray band). The measurement result from this work is $1.868 \pm 0.006$ b (closed circle).
  }
  \label{fig_nsigma}
\end{figure}

The measurement results for gas I and gas I\hspace{-.1em}I calculated using Eq.~\ref{eqNsigma} with $\sigma(^{3}{\mathrm He})=5333\pm7$ b~\cite{Mughabghab2006} are shown in Table~\ref{tab_result}.
We obtained a combined value for $\sigma({}^{14}{\rm N})$ as $1.868\pm 0.003$ (stat.) $\pm 0.006$ (sys.) b, with an accuracy of 0.4\%.
The values for $\sigma({}^{14}{\rm N})$ obtained in this work as well as from the previously published work are shown in Fig.~\ref{fig_nsigma} for a comparison.
Our result is consistent with the weighted average (= $1.84 \pm 0.03$ b) calculated from the previously published results.

As a result of our measurement, the improved accuracy in calculating the cross section ratio of $\sigma({}^{14}{\rm N})/\sigma({}^{3}{\rm He})$ is reduced $1/5$ times less than the ratio of recommended values~\cite{Mughabghab2006}.
Such improvement was attained by minimizing the uncertainty in the ${}^{3}{\rm He}$ content ratio in G1He gas.
Additionally, the amount of ${}^{3}{\rm He}$ contained in G1He gas is nominally 10\% of the admixed ${}^{3}{\rm He}$ that is introduced into the TPC under current experimental conditions.
Consequently, the resulting uncertainty of $\rho({}^{3}{\rm He})$ becomes approximately 0.04\%.
Therefore, we conclude that our experimental results show an improvement in the accuracy of neutron lifetime calculation by significantly reducing the sources of uncertainties in $\rho({}^{3}{\rm He})$.
\begin{table}[htp]
\caption{Experimental results}
\begin{center}
\begin{tabular}{c c c}\hline
parameter & Gas I & Gas I\hspace{-.1em}I \\\hline\hline
$\rho(^3 \mathrm{He})/\rho(^{14} \mathrm{N})$ & $2.351(7) \times 10^{-4}$ & $5.212(12) \times 10^{-4}$ \\
$\varepsilon(^{3} \mathrm{He})/\varepsilon(^{14} \mathrm{N})$ & $1.0007 \pm 0.0007$ & $1.0007 \pm 0.0007$ \\
& 1.4915 & 0.6712\\
$S(^{14} \mathrm{N})/S(^3 \mathrm{He})$&$\pm 0.0044$ (stat.)&$\pm 0.0011$ (stat.)\\
&$\pm 0.0013$ (sys.)&$\pm 0.0007$ (sys.) \\\hline
& 1.871& 1.867 \\
$\sigma(^{14}\mathrm{N})$ (b)&$\pm 0.005$ (stat.)&$\pm 0.003$ (stat.)\\
&$\pm 0.006$ (sys.)&$^{+0.005}_{-0.006}$ (sys.)\\\hline\hline
Total $\sigma(^{14}\mathrm{N})$ (b)& \multicolumn{2}{c}{$1.868 \pm 0.003$ (stat.) $\pm 0.006$ (sys.)}\\\hline
\end{tabular}
\end{center}
\label{tab_result}
\end{table}
\clearpage
\section{DISCUSSION}
\label{discussion}
\begin{figure}[th]
\centering
  \includegraphics[width=8cm]{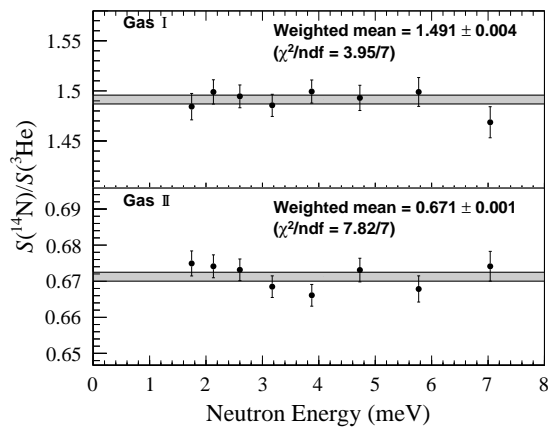} 
  \caption{$S(^{14}\mathrm{N})/S(^{3}\mathrm{He})$ of each of the bunched neutron beams in gas I (up) and gas I\hspace{-.1em}I (low). The gray band indicates the weighted mean and the scaled error. }
  \label{fig_eventratio}
\end{figure}

The present results deduced by Eq.~\ref{eqNsigma} are on the assumption of the 1/$v$ law.
The validity of the 1/$v$ law in our energy region, 1--7 meV, is necessary to be discussed.
In general expression for the reaction cross section in low neutron energy region, the reduced cross section, $\sigma(E)\sqrt{E}$ where $E$ is a neutron energy is expressed as,
\begin{eqnarray}
\label{eqvinverse_define}
\sigma(E)\sqrt{E}=a+b\sqrt{E},
\end{eqnarray}
where $a$ and $b$ are free parameters~\cite{Keith2004}.
According to the previous measurements from thermal energy to a few eV~\cite{Koehler1989,Koehler1993,Keith2004}, the dispersions of $^{3}$He(n,p)$^{3}$H and the $^{14}$N(n,p)$^{14}$C reactions from the 1/$v$ law, $b\sqrt{E}/a$ are estimated as less than 0.02\% and 0.05\% below 25.3 meV, respectively.
The event rate ratios of $S(^{14}\mathrm{N})/S(^{3}\mathrm{He})$ of neutron bunches corresponding to the average neutron energy for gas I and gas I\hspace{-.1em}I are shown in Fig.~\ref{fig_eventratio}, and did not show any dependence on the neutron energy with $O$(0.1\%) accuracy, as they were expected. Our result of $\sigma({^{14}\mathrm{N}})=1.868(6)$ b was also extrapolated to the thermal neutron energy of 25.3 meV from 1--7 meV. The dispersion, which is at most 0.05\% from the above discussion, was negligibly small.

The thermal cross section of the $^{14}$N(n,p)$^{14}$C reaction obtained in this work, $1.868 \pm 0.006$ b, is the most accurate value reported in the literature thus far. 
The improved accuracy achieved in this work is due to the accurate evaluation of the number density ratio using the gas handling system, which resulted in five times smaller uncertainty in our result as compared to the previously reported values.
The number density measurements of $\rho({}^{3}{\rm He})$ by both---the DM and VE methods---showed similar values within the uncertainty, which corroborates the reliability of the VE method that is also used for the neutron lifetime experiment.
The main uncertainty in our measurements was derived from the number density of the $^{3}$He nuclei, which is also a subject of concern for the neutron lifetime measurement. 
The number density values may be evaluated with better accuracy using the VE method with a renewed gas handling system in future work, and this may also improve the accuracy of $\sigma({}^{14}{\rm N})$.

The accurate determination of the $^{14}$N(n,p)$^{14}$C cross section is valuable in various fields.
For instance, the reaction caused in the atmosphere is known to produce $^{14}$C, that is used for $^{14}$C dating~\cite{Currie2004,Enoto2017}.
This element is also produced as a remarkable product in the atomic reactors \cite{Sohn2003}.
Additionally, in boron neutron capture therapy, this is one of the main reactions that produce high linear energy transfer protons, when thermal neutrons are injected into the human body~\cite{Suzuki2014}.
The cross section values at the cosmological temperatures can be used for estimating the amount of isotopes produced in the slow-neutron capture processes in asymptotic giant branch stars, which can be compared with the experimentally measured values~\cite{Brehm1988,Koehler1989,Sanami1997,Kii1999,Wallner2016}.
The value of $\sigma({}^{14}{\rm N})$ is often used for the analysis of other reactions and evaluating the neutron flux \cite{Issa2017}.
For instance, the cross section of the $^{17}$O(n,$\alpha$)$^{14}$C reaction at $v_0$ was determined relative to that of the $^{14}$N(n,p)$^{14}$C reaction~\cite{17OWagemans1998,14NWagemans2000}.
The reaction cross section value of $257 \pm 10$ mb was determined by Wagemans {\it et al.}, using $\sigma(^{14}\mathrm{N})=1.93 \pm 0.05$ b \cite{14NWagemans2000}.
We redetermined the cross section value as $249 \pm 6$ mb using our results of $\sigma(^{14}\mathrm{N})$ obtained in this work.

In principle, this measurement method can be extended to other (n,p) or (n,$\alpha$) reactions that utilize a gas target.
Furthermore, a list of plausible reactions involving gas targets, including the $^{17}$O(n,$\alpha$)$^{14}$C reaction, is presented in Table~\ref{tab_reactions}, in which the proposed method might be utilized for the evaluation of the reaction cross sections.
\begin{table}[htp]
\caption{List of cross sections of the (n,p) and (n,$\alpha$) reactions at $v_0$ from gas targets. The recommended values of the cross sections~\cite{Mughabghab2006} and the experimental results\cite{Lal1987,Jiang1990} are presented.}
\begin{center}
\begin{tabular}{c c c c c} \hline 
Reaction &Q-value (keV)&Cross Section (b)&Unc. (\%)&Available gas \\\hline\hline
$^{3}$He(n,p)$^{3}$H        &764 &5333(7) \cite{Mughabghab2006}& 0.13& He\\
$^{10}$B(n,p)$^{10}$Be      &226 &$6.8(5) \times 10^{-3}$ \cite{Lal1987}& 7.4& BF$_{3}$ \\
$^{10}$B(n,$\alpha$)$^{7}$Li&2790&3837(9) \cite{Mughabghab2006}& 0.23& BF$_{3}$ \\ 
$^{14}$N(n,p)$^{14}$N&626&1.86(3) \cite{Mughabghab2006}&1.6& N$_{2}$ \\ 
$^{17}$O(n,$\alpha$)$^{14}$C&1818&$235(10) \times 10^{-3}$ \cite{Mughabghab2006}&4.3& CO$_{2}$ \\ 
$^{33}$S(n,p)$^{33}$P&534&$2(1) \times 10^{-3}$ \cite{Mughabghab2006}& 50& SF$_{6}$ \\ 
$^{33}$S(n,$\alpha$)$^{30}$Si&3494&0.115(10) \cite{Mughabghab2006}& 42& SF$_{6}$ \\
$^{36}$Ar(n,p)$^{36}$S&73& $<$ 1.5 $\times 10^{-3}$ \cite{Jiang1990}&-& Ar \\ 
$^{36}$Ar(n,$\alpha$)$^{33}$S&2001&$5.5(1) \times 10^{-3}$ \cite{Mughabghab2006}& 1.8& Ar \\\hline
\end{tabular}
\end{center}
\label{tab_reactions}
\end{table}

\section*{ACKNOWLEDGE}
This work was supported by JSPS KAKENHI Grant Number JP19GS0210, JP26247035, JP23244047, JP20340051, JP24654058, JP16H02194 and JP16J10830.

The neutron experiment at the Materials and Life Science Experimental Facility of the J-PARC was performed under a user program (Proposal No. 2014B0271) and S-type project of KEK (Proposal No. 2014S03).

\bibliographystyle{unsrt}

\end{document}